\documentclass[12pt]{article}
\usepackage[dvips]{graphicx}
\begin{document}
\begin{center}   {SHM of Galaxies Embedded within Condensed Neutrino Matter}
\end{center}
\vspace{.2in}
\begin{center}
P.D. Morley \\  
System of Systems Analytics, Inc. \\ 11250 Waples Mill Road \\ Suite 300 \\Fairfax, VA 22030-7400 \\
{\it E-mail address:} peter3@uchicago.edu \\ and\\
D.J. Buettner \\
Aerospace Corporation\\P. O. Box 92957\\Los Angeles, CA 90009-2957\\{\it E-mail address:} Douglas.J.Buettner@aero.org
\end{center}
\begin{abstract}
We re-examine the question of condensed neutrino objects (degenerate neutrino matter) based on new calculations. The potential show-stopper issue of free-streaming light neutrinos inhibiting galaxy formation is addressed. We compute the period associated with simple harmonic motion (SHM) of galaxies embedded within condensed neutrino objects. For observational consequences, we examine the rotational velocities of embedded galaxies using Hickson 88A (N6978) as the prototype.  Finally, we point out that degenerate neutrino objects repel each other in overlap and we compute directly the repulsive force between two interesting and relevant configurations. An outstanding issue is whether the accompanying tidal forces generated by condensed neutrino matter on embedded galaxies give rise to galactic bulges and halos.
\end{abstract}
keywords: Galaxies: bulge - Galaxies: interactions - Equation of state 

\newpage
\section{Introduction}
There are two known astrophysical objects containing degenerate matter supporting equilibrium: white dwarfs and neutron stars. A possible third degenerate regime was investigated decades ago$^{1}$, hypothetically lying in between a neutron star and a black hole, but the equation of state required was so unnatural$^{1}$, it is believed that `quark stars' cannot exist. 

Another possible third degenerate regime might be found in galactic halos. However, the required {\bf multi-kilo-parsec hydrostatic equilibrium configuration} means kinetically that the particle must have substantially less mass than the electron itself\footnote{Heavy particles like neutrons give rise to Manhattan-size equilibrium configurations, while light particles like electrons give rise to earth-size equilibrium configuration. To obtain kilo-parsec equilibrium configurations requires a particle substantially lighter than the electron.}. The only known particles are the neutrinos. In this paper, we discuss this possible third degenerate regime: Dirac neutrino matter consisting primarily of big bang relic neutrinos that have condensed from the primordial neutrinos and anti-neutrinos of the Big Bang itself.

The history of degenerate neutrino matter starts in 1973 with the proposal of Cowsik and McClelland$^{2}$ that dark galactic halos may be neutrinos. There followed an intense period of research into this idea, up to the mid-1980's, until it was realized that light neutrinos will have large velocity dispersions making condensation by gravitational relaxation unlikely. A very good review of the literature of the 1980's can be obtained from the work of Chubb$^{3}$ who cites Gao and Ruffin$^{4}$, Fabbri et al$^{5}$, Zhang et a$^{6}$ for degenerate galactic neutrino halos and Tremaine and Gunn$^{7}$, Schramm and Steigman$^{8}$, Ruffini and Stella$^{9}$, Bond et al$^{10}$, Melott$^{11, 12}$ and Madsen and Epstein$^{13}$ for semi-degenerate galactic matter halos. More recent work$^{14}$ states that a cold relic neutrino gas leads to the Lane-Emden equation.  In contrast, we derive our EOS for the condensation of big bang relic neutrinos using the most recent particle data available$^{15}$ indicating that neutrinos have mass of the order of $\sim$ 1 eV/c$^{2}$ (c: speed of light), with very small mass differences between neutrino flavors. Thus, with a common neutrino mass and a mass degeneracy of 6, we obtain an EOS that does not lead to the Lane-Emden equation. In our present solution, we assume no differential rotation of the neutrino object. The resulting no rotation degenerate neutrino objects have size and mass which, if they exist, are the most massive and largest single objects in the universe. Their size ranges from objects that can contain a single galaxy to objects that can easily contain a cluster of galaxies. Furthermore, they are totally transparent due to the non-interacting nature of low energy neutrinos: one can see right through them. Light, however, does carry information that will reveal their existence. We reference this observational property in the conclusion section of this paper.  We show, by deriving the theoretical formula, that a galaxy near the center of condensed neutrino matter undergoes simple harmonic motion (SHM). This spatial separation of galactic mass centroids with degenerate neutrino object mass centroids can lead to asymmetric galactic rotation velocity profiles (meaning the galaxy rotation speeds have azimuth angle dependence). We show how this works using Hickson 88A as the iconic example. Finally, we comment on the expected tidal forces that will arise for galaxies embedded within these neutrino objects.

\section{A Cooling Mechanism for Primordial Neutrinos and Anti-neutrinos}
As already mentioned, light neutrinos pose a cosmological show-stopper$^{16}$: their free-streaming relativistic velocities wash-out small-scale primordial density fluctuations, which would have allowed galaxies to form before galaxy clusters. If, however, neutrino condensation occurs and occurs early on, then the neutrinos are swept up and removed as a gas, negating the free-streaming phase. The question now becomes: what is the cooling mechanism for cosmological neutrinos and anti-neutrinos that would allow them to lose their kinetic energy around the time they decouple from baryonic matter?

A possible cooling mechanism has been given in ref 17, where it was shown, in a classical physics computation, that neutrinos radiate power through their anomalous magnetic dipole moment in a turbulent magnetic field. The early universe had magnetic fields comparable to or larger than those in Neutron stars$^{18,19}$. In the absence of a mean field, all magnetic fields are generated and maintained by the turbulence itself. Furthermore, all turbulent magnetic fields are out of thermal equilibrium$^{20}$. A small-scale dynamo gives exponentially growing magnetic fields with energy concentrated at small scales. If the (Dirac) neutrino magnetic moment is anomalous, it can have values approaching the Particle Data Group$^{15}$ upper bound 
\begin{equation}
\mu_{\nu} < 0.32 \times 10^{-10} \mu_{B}
\end{equation}
where $\mu_{B}$ is the Bohr magneton. There are regions in early universe phase space$^{17}$ where neutrino magnetic moments can radiate power at 1 MeV/s. Comparing this to decoupling neutrino temperatures $\sim 1-3$ MeV, neutrino and anti-neutrino cooling may be realizable. In order to answer the question of neutrino condensation timelines, one needs to do a full electromagnetic treatment of early universe magnetic field generations, in concert with neutrino primordial density fluctuations. For this paper, we assume that magnetic cooling in turbulent early-universe plasmas, which are known to be present at the time of neutrino and anti-neutrino decoupling from baryonic matter, is sufficient for early condensation of neutrino objects. This then would be a third example of degeneracy pressure supporting an astrophysical body.

\section{Scattering Processes within the Neutrino Object}
The neutrino object proper will be constructed in section 4. Before that, we discuss its transparency, internal interactions and lifetime. 
\subsection{Photon scattering and neutrino-photon annihilation in condensed neutrino matter}
Ref 21 has computed the low energy neutrino-photon scattering in vacuum (the box diagram in the Standard Model). Taking his numerical calculations, we graph the cross section in Figure 1, for visible light. It is arguably one of the smallest cross sections in nature. However, the scattering in condensed neutrino matter is taking place within a degenerate medium and so there are no empty neutrino states available except for those near the Fermi level. Thus the true incoming photon cross section is dramatically much smaller than Figure 1 in vacuum and we conclude that the condensed neutrino matter is totally transparent to light, except scattering by embedded matter within it.

\begin{figure}[htp]
\includegraphics[scale=0.75]{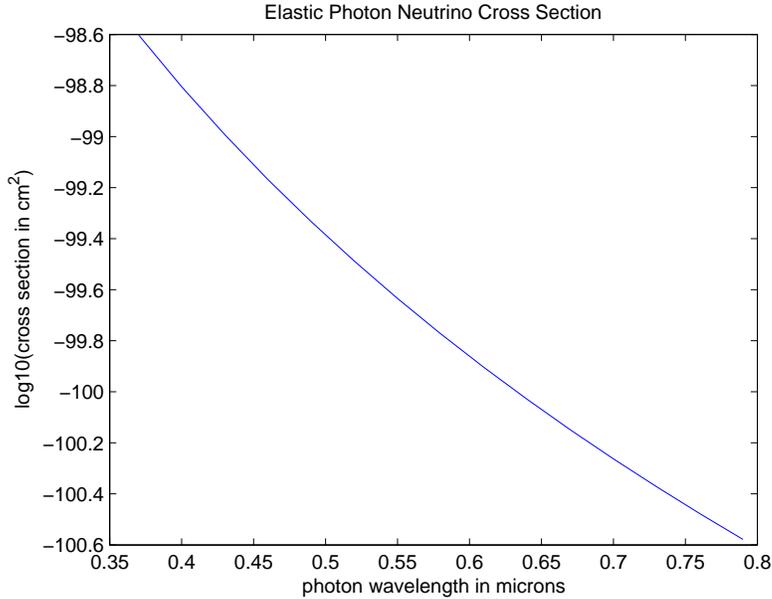}
\caption[*]{Elastic photon-neutrino scattering in vacuum$^{21}$.}
\label{Fig. 1}
\end{figure}

The Dicus calculation$^{21}$ also gives us the annihilation cross section (the inverse reaction) of neutrino-anti-neutrino to photons in condensed neutrino matter. This rate is so small that condensed neutrino matter has a lifetime that is orders of magnitude longer than the lifetime of the universe. Once formed, their only evolution is to grow and potentially collapse to Black Holes.

\subsection{Neutrino-anti-neutrino annihilation into electron and positron}
Without an exact calculation, one would think that the critical mass (maximum allowable mass) of condensed neutrino matter is determined by the neutrino-anti-neutrino annihilation into an electron and positron, Figure 2, in analogy with the maximum white dwarf star being determined by the disappearance of electrons (the source of pressure resisting gravity) from inverse beta-decay. This turns out not to be the case. The condensed neutrino object critical mass and size come directly from the requirement that its radius must be greater than its Schwarzschild's radius. This maximum condensed neutrino mass turns out to have a Fermi momentum much smaller than the $m_{e}c$ (where $m_{e}$ is the electronic mass) value coming from Figure 2, and so Figure 2 is absent due to conservation of energy.

\begin{figure}[htp]
\includegraphics[scale=0.75]{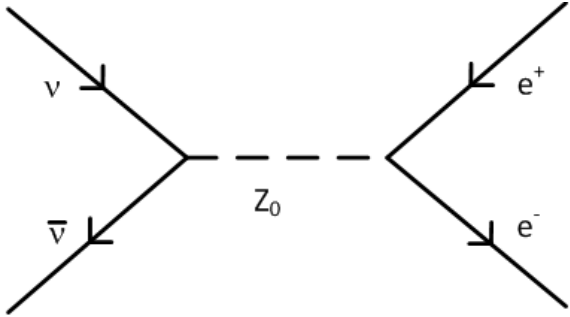}
\caption[*]{$\nu - \overline{\nu}$ annihilation into $e - \overline{e}$ (Forbidden by energy conservation).}
\label{Fig. 2}
\end{figure}

\subsection{Neutrino Dipole Interactions}
Within the condensed object, neutrinos and anti-neutrinos will have magnetic dipole interactions. Because the degenerate state has configuration states that are plane waves, only the Fermi Contact Interaction contributes. It is easy to show that this interaction energy vanishes as the size of the object becomes large, and so neutrinos and anti-neutrinos are ideal degenerate gases.

\section{Equation of state}
To simplify the description of degenerate neutrino objects, we use the fact that the experimental masses are close in value to each other$^{15}$, $\sim$ 1 eV/c$^{2}$. We call this the `common neutrino mass'. The various configurations possible with three flavors of Dirac neutrinos and anti-neutrinos can be simplified by assuming that neutrino objects have no net leptonic charge so the total degeneracy factor is 6 (three flavors of Dirac neutrino and three flavors of Dirac anti-neutrinos). The condensation process is non-trivial as already discussed and will put constraints on the value of the neutrino anomalous moment and early universe magnetic turbulence. 

In white dwarfs, electrons constitute the pressure, but contribute negligible mass. The mass in white dwarfs comes from nuclei which must be present to preserve charge neutrality. Here, the situation is totally different because the degenerate neutrino gas supplies both the pressure and the mass. 

The number of neutrinos (spin up and spin down) between momentum states p and p + dp is (restoring c, $\hbar$ units)
\begin{equation}
N(p)dp = \frac{V8 \pi p^{2}}{h^{3}}dp
\end{equation}
where $V$ is the volume and $h$ is Planck's constant. The total neutrino particle energy is $E_{\nu}$
\begin{equation}
E_{\nu} = \int_{0}^{p_{0}}N(p)E(p)dp
\end{equation}
where $p_{0}$ is the Fermi level at radius $r$ and
\begin{equation}
E(p) =  \sqrt{m_{\nu}^{2}c^{4}+p^{2}c^{2}} \; .
\end{equation}
Once we obtain $E_{\nu}$ as a function of $p_{0}$, the mass density $\rho_{\nu}$ as a function of $p_{0}$ is
\begin{equation}
\rho_{\nu} = \frac{E_{\nu}}{Vc^{2}} \; .
\end{equation}
The reason for this is that the neutrinos are in momentum states and their mass increases in accordance with Special Relativity. Eq.(3) can be integrated by the change of variables
\begin{equation}
\sinh \theta = \frac{p}{m_{\nu}c}
\end{equation}
giving
\begin{equation}
\rho_{\nu} = \frac{\pi m_{\nu}^{4}c^{3}}{h^{3}} \{ \frac{1}{4} \sinh 4 \theta - \theta \}|_{\theta =\theta_{0}}
\end{equation}
Using
\begin{equation}
\sinh 4 \theta = 2(2 \sinh \theta \cosh \theta)(2 \cosh^{2} \theta -1)
\end{equation}
and changing to the $x$ variable
\begin{equation}
x = \frac{p_{0}}{m_{\nu}c}
\end{equation}
we obtain
\begin{equation}
\rho_{\nu} = \frac{\pi m_{\nu}^{4}c^{3}}{h^{3}} \{2x(1+x^{2})^{\frac{3}{2}}-x\sqrt{1+x^{2}}-\sinh^{-1} x \} \; .
\end{equation}
The total mass density of the 6 different neutrinos and anti-neutrinos making up condensed neutrino matter is $\rho_{total}$
\begin{equation}
\rho_{total} = \frac{6 \pi m_{\nu}^{4}c^{3}}{h^{3}} \{2x(1+x^{2})^{\frac{3}{2}}-x\sqrt{1+x^{2}}-\sinh^{-1} x \} \; .
\end{equation}
The total pressure $P_{total}$ is just 6$\times$ the pressure of the familiar electron gas giving
\begin{equation}
P_{total} = \frac{2 \pi m_{\nu}^{4}c^{5}}{h^{3}} \{ x(2x^{2}-3)\sqrt{1+x^{2}}+3 \sinh^{-1} x \} \; .
\end{equation}
Eq(11) and Eq(12) are the equation of state (EOS) for a degenerate neutrino gas with 6 degrees of flavor degeneracy. It does not result in the Lane-Emden equation that appears in white dwarfs$^{22}$.
\section{Hydrostatic equilibrium}
The equation of hydrostatic equilibrium to be solved is$^{22}$
\begin{equation}
\frac{1}{r^{2}}\frac{d}{dr}(\frac{r^{2}}{\rho_{total}}\frac{dP_{total}}{dr}) = -4\pi G \rho_{total} \; .
\end{equation}
Eq(13) with an iron-silicate EOS describes earth, with an electron EOS describes white dwarfs, and with a neutrino EOS describes condensed neutrino matter, if it exists. In what must be considered one of the most unusual results in physics, the smallest particle in mass gives rise to the largest and most massive structures in the universe. 

The boundary conditions associated with Eq(13) are
\begin{eqnarray}
x & = & x_{0}, \; @ r = 0 \nonumber \\
\frac{dx}{dr} & = & 0, \; @ r = 0
\end{eqnarray}
The hydrostatic equation is integrated outward to a radius $R_{0}$ where $x$ = 0. The mass interior to a specified point $r \leq R_{0}$ is
\begin{equation}
M(r) = 4 \pi \int_{0}^{r} \rho_{total} r^{2} dr
\end{equation}
with total mass $M_{total} = M(R_{0})$. The total potential energy $\Omega$ of condensed neutrino matter is
\begin{equation}
\Omega = -G \int_{0}^{R_{0}} \frac{M(r)}{r} 4 \pi r^{2} \rho_{total}(r) dr \; .
\end{equation}
The astrophysical characteristics are
\begin{enumerate}
\item Total mass $M(R_{0})$ and radius $R_{0}$.
\item Density profile $\rho_{total}(r)$.
\item Mass profile $M(r)$.
\item Potential energy $\Omega$.
\end{enumerate}
\subsection{Spherical Equilibrium Solutions}
We do the cases $m_{\nu}c^{2}$ = 1 eV and the $m_{\nu}c^{2}$ = 0.5 eV. The abbreviation `ly' stands for light year, `pc' stands for parsec and the abbreviation $R_{S}$ means Schwarzschild radius.

 \begin{table}[h]
\begin{center}
     \begin{tabular}{ccccc}
     \multicolumn{5}{c}{ $m_{\nu}c^{2}$ = 1 eV} \\  \hline
       $x_{0}$ & $M_{total}$ in M$_{\odot}$ & $R_{0}$ (in ly) & $\Omega$ (in Joules) & $R_{0}$ (in $R_{S}$) \\ \hline
       \mbox{} & \mbox{} & \mbox{} & \mbox{} & \mbox{} \\
  0.001 & $3.100\times10^{13}$ & $13.01\times10^{6}$ & $-1.764\times10^{54} $ & $1.344 \times 10^{6}$ \\
   0.01  &     $9.809\times10^{14}$  &  $4.12\times10^{6}$  &  $-5.569\times10^{57}$ & $1.345 \times 10^{4}$ \\
   0.1   &     $3.077\times10^{16}$  &  $1.30\times10^{6}$  &  $-1.739\times10^{61}$ & 135.3 \\
   1.0  &  $5.27\times10^{17}$ &  $3.71\times10^{5}$  & $-1.969\times10^{64}$  &  2.254 
      \end{tabular}
\end{center}
      \caption{Hydrostatic solutions for degenerate neutrino matter}
      \end{table}

From Table 1, it is clear that the transparent condensed neutrino matter can contain one or more galaxies embedded within them. The case $x = 1$ is near the applicability of Eq.(13), since its solution has a radius $\sim 2.25 \times$ its Schwarzschild's radius. Thus it is manifestly clear that neutrino-anti-neutrino annihilation into electrons and positrons never occurs: Figure 2 is forbidden by conservation of energy until the Fermi level is high enough, but by then, condensed neutrino matter will have already collapsed under its own weight. In the following figures 3-6, we give the computed solutions. In Table 2, we give the units for the figures 3-5 ($m_{\nu}c^{2}$ = 1 eV). As the neutrino mass decreases, condensed neutrino matter becomes more massive and larger for the same boundary condition $x(0)$ at the center.

\begin{table}[h]
\begin{center}
     \begin{tabular}{ccc}
     \multicolumn{3}{c}{ $m_{\nu}c^{2}$ = 1 eV} \\  \hline
       mass unit (M$_{\odot}$) & distance unit (ly) & density unit (gm/cm$^{3}$) \\ \hline
       \mbox{} & \mbox{} & \mbox{} \\
   3.213$\times10^{17}$    &   1.5045$\times10^{5}$  &  1.76307$\times10^{-20}$  \\
      \end{tabular}
\end{center}
      \caption{Units for Figures 3-5}
      \end{table}

\begin{figure}[htp]
\includegraphics[scale=0.75]{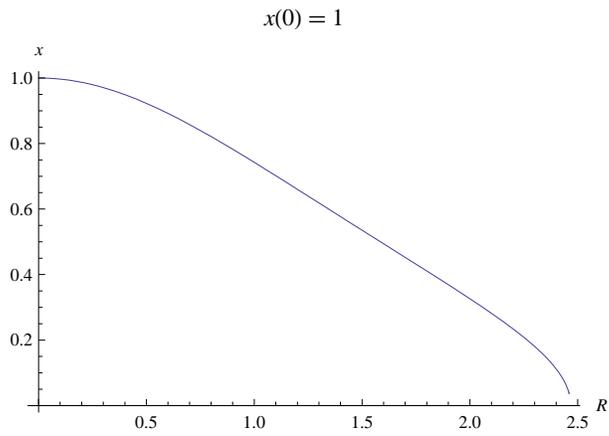}
\caption[*]{Variation of the reduced Fermi momentum $x$ for x(0) = 1.}
\label{Fig. 3}
\end{figure}

\begin{figure}[htp]
\includegraphics[scale=0.75]{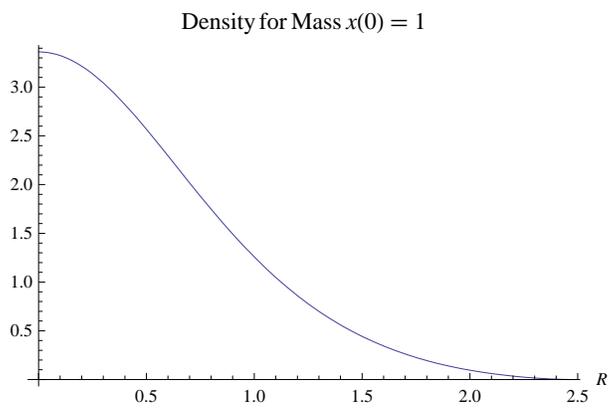}
\caption[*]{Density variation for x(0) = 1.}
\label{Fig. 4}
\end{figure}

\begin{figure}[htp]
\includegraphics[scale=0.75]{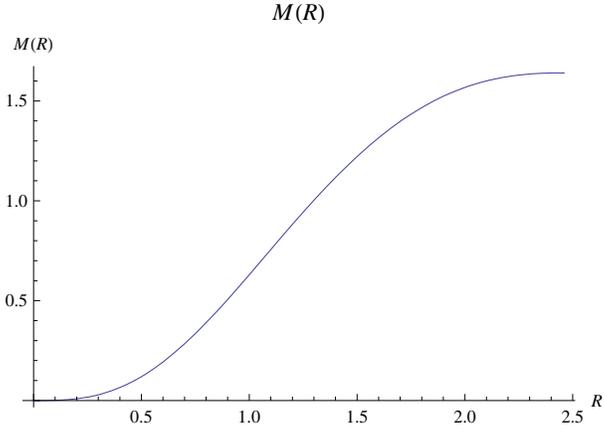}
\caption[*]{Integrated mass for x(0) = 1.}
\label{Fig. 5}
\end{figure}

\begin{figure}[htp]
\includegraphics[scale=0.75]{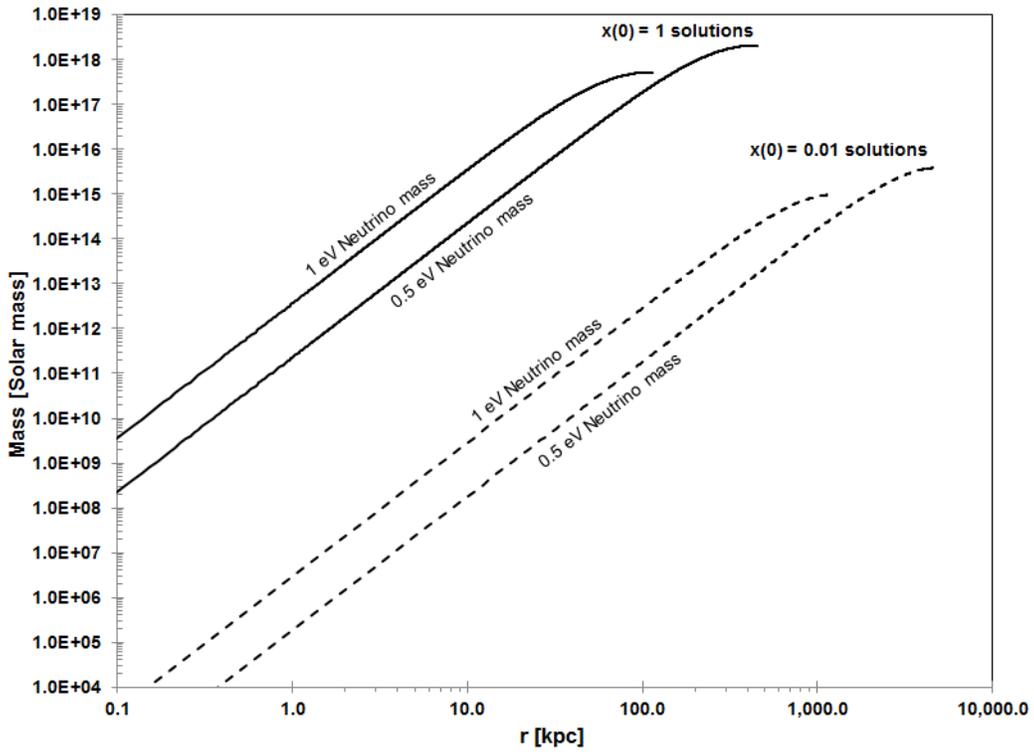}
\caption[*]{Condensed neutrino objects, mass and radii distributions.}
\label{Fig. 6}
\end{figure}

\subsection{Non-spherical Solutions}
Condensed neutrino matter may support non-spherical solutions, because of the absence of surface tension forces. Bi-spherical geometry with a bridge or filament in-between two nearly spherical ends, giving a dumbbell appearance would be attractive first-approach. This, however, will not be pursued further here. 

\section{The Repulsive Force Between Condensed Neutrino Objects}
Another important concept which the existence of the static neutrino solutions affords us is the calculation of the repulsive force between overlapping neutrino objects. If two objects overlap, then their neutrino wavefunctions overlap, which causes them to disappear under the anti-symmetric permutation operator. They then have to appear in the available un-occupied states. Thus, as two degenerate  condensed neutrino objects are pushed together, the radius of the larger object decreases with concomitant increasing of its Fermi level. The amount of energy required to push two objects together with an overlap of distance $L$ is thus equal to the difference of energies between the original larger in size object and the newly compressed object whose radius has decreased by length $L$. The new `binary' configuration now has both objects just-touching.  As an example of the immense repulsion forces involved, we compute the repulsive force between the x(0) = 0.1 object and a x(0) = 0.05 object in Figure 7. The units in this Figure are the same units as in Table 2.  Thus, for example, a distance $R$ =1 is an overlap distance of  1.5045$\times10^{5}$ light years and the energy unit is the mass unit of Table 2 multiplied by c$^{2}$. The original size of these objects, in distance units is 8.65183 for the higher mass object x(0) = 0.1 and 12.24828 for the smaller mass object x(0) = 0.05. The actual force is the slope of the energy curve in Figure 7, which to the numerical accuracy used in the calculation, is parabolic. Thus the repulsive force is a compressed spring, a linear force. Working out the units, we use the symbol $k^{.1}_{.05}$ to designate the spring constant, which has the value
\begin{equation}
k^{.1}_{.05} = 3.00 \times 10^{20} \;   kg/s^{2} \; .
\end{equation}
This is a candidate for the largest spring constant in Nature. From these considerations, multiple degenerate neutrino objects may be found experimentally packed together like billiard balls. This has observational implications: the space between the "billiard balls" may be the accumulation region for baryonic matter and the prediction that matter in the early universe congregates in ribbons and the potential wells at the center of neutrino objects.

\begin{figure}[htp]
\includegraphics[scale=0.75]{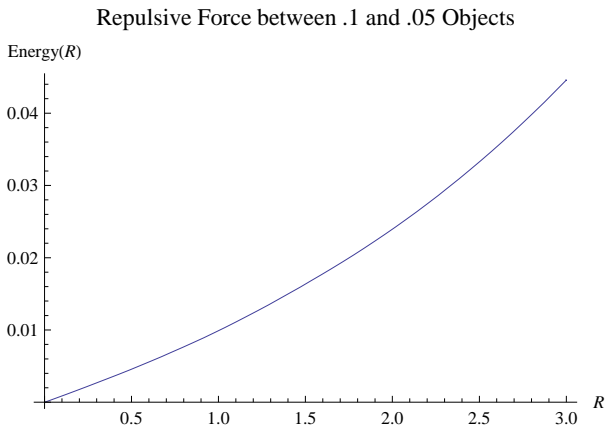}
\caption[*]{The Fermi repulsion potential energy between overlapping x(0) = 0.1 and x(0) = 0.05 objects.}
\label{Fig. 7}
\end{figure}

\section{Simple harmonic motion of an embedded galaxy}
In nearly four hundred years of celestial mechanics we have only seen bound gravitational systems executing elliptical orbits about their common center of mass and unbound systems executing parabolic trajectories. However, the dynamical problem of a galaxy embedded within condensed neutrino matter is unprecedented: there are no foci. Condensed neutrino matter in hydrostatic equilibrium is spherically symmetric, so the force vanishes at the center. As an embedded galaxy moves past the center, gravity will act to pull it back: the galaxy will execute simple harmonic motion (SHM).  In this section, we derive the period of this motion.

The total interaction $W_{total}$ between the embedded galaxy and the condensed neutrino matter is
\begin{equation}
W_{total} = \frac{-G}{2} \int \int \frac{\rho(\vec{r})\rho(\vec{r}^{\prime})}{|\vec{r} - \vec{r}^{\prime}|} d^{3}rd^{3}r^{\prime}
\end{equation}
where $\rho$ is the total mass density of the system. Separating out the condensed neutrino matter `B' from the embedded galaxy `G' by $\rho = \rho_{B} +\rho_{G}$ gives
\begin{equation}
W_{total} = {\rm self \; energy} -G \int \int \frac{\rho_{G}(\vec{r})\rho_{B}(\vec{r}^{\prime})}{|\vec{r} - \vec{r}^{\prime}|} d^{3}rd^{3}r^{\prime}
\end{equation}
Defining $W_{interaction}$ as
\begin{eqnarray}
W_{interaction} & = & -G \int \int \frac{\rho_{G}(\vec{r})\rho_{B}(\vec{r}^{\prime})}{|\vec{r} - \vec{r}^{\prime}|} d^{3}rd^{3}r^{\prime} \nonumber \\
& \equiv & \int \rho_{G}(\vec{r}) \Phi(\vec{r}) d^{3}r 
\end{eqnarray}
with
\begin{equation}
\Phi(\vec{r})  = -G \int \frac{\rho_{B}(\vec{r}^{\prime})}{|\vec{r} - \vec{r}^{\prime}|}d^{3}r^{\prime} \; .
\end{equation}
We next expand $\Phi(\vec{r})$ in a Taylor series expansion about the origin using the general formula
\begin{equation}
f(\vec{r}+\vec{a}) = e^{\vec{a} \cdot \vec{\nabla}}f(\vec{r})
\end{equation}
In our case
\begin{eqnarray}
\Phi(0) & = & 0 \nonumber \\
\vec{\nabla} \Phi(\vec{r})|_{\vec{r}= 0} & = & \vec{0}
\end{eqnarray}
and the interaction is quadratic in the small amplitude $a$ signaling SHM. Computing the non-zero term gives
\begin{equation}
\Phi(\vec{r}) \rightarrow \int \frac{\rho_{B}(\vec{r}^{\prime})a^2}{|\vec{r}^{\prime}|^{3}} d^{3}r^{\prime}
\end{equation}
finally leading to
\begin{equation}
W_{interaction} = -Ga^{2}M_{G} \int \frac{\rho_{B}(\vec{r}^{\prime})}{|\vec{r}^{\prime}|^{3}} d^{3}r^{\prime} \; .
\end{equation}
In Eq(25), $M_{G}$ is the mass of the galaxy executing SHM near the center of the condensed neutrino matter. Recall from classical physics that the work done by a spring force for displacement $a$ is $W_{interaction} = -\frac{1}{2}ka^{2}$, we see then that the angular frequency $\omega$ associated with motion about the condensed neutrino matter centroid is
\begin{equation}
\omega = \sqrt{2G \int \frac{\rho_{B}(\vec{r}^{\prime})}{|\vec{r}^{\prime}|^{3}} d^{3}r^{\prime} }
\end{equation}
which leads to the final formula for the period $T$
\begin{equation}
T = \left( \frac{2G}{\pi} \int \frac{\rho_{B}(r^{\prime})}{r^{\prime}}dr^{\prime} \right)^{-1/2}
\end{equation}
In Table 3, we present the period for two interesting cases. Condensed neutrino matter has effectively two parameters: the common neutrino mass and the boundary condition $x(0)$ of the centroid Fermi momentum. This parameter space creates SHM periods ranging from millions to billions of years.

\begin{table}[h]
\begin{center}
     \begin{tabular}{cc|c}
     \multicolumn{2}{c}{$m_{\nu}c^{2}$ = 1 eV} & \multicolumn{1}{c}{$m_{\nu}c^{2}$ = 0.5 eV}  \\  \hline
       x(0) & Period (years) & Period (years)  \\ \hline
       \mbox{} & \mbox{}  & \mbox{}  \\
   0.01   &   208$\times10^{6}$ & 832$\times10^{6}$ \\
    0.1   &   6.93$\times10^{6}$ & 27.72$\times10^{6}$ \\
      \end{tabular}
\end{center}
      \caption{Period of SHM for selected condensed neutrino matter objects}

      \end{table}

\section{Expected observables of SHM}
For a galaxy undergoing SHM, we expect to see clear separation between the center of visible mass and the gravitational potential. But the most interesting consequence lies in the expected tidal forces that will act on the embedded galaxy. These tidal forces change direction as SHM is executed, which means that a detailed simulation is required to answer quantitatively the cumulative effect of many oscillations. This is research beyond the topic of the present paper. However, one expects that stellar orbits become randomized due to this `tidal friction'. The possible outcome might be the formation of a galactic bulge and halo. Presently, the Milky Way galactic bulge is thought to occur from two scenarios$^{23}$: secular evolution of the galactic disk (i.\ e. disk instability) and/or gravitational collapse. However, both would be expected to be consequences of tidal forces: the distinction arising from the difference of the initial local star density resisting the tidal gradient. A very interesting number then is the time it would take for an embedded galaxy to thoroughly randomize all stellar orbits and create an elliptical galaxy.

The predicted spatial separation of galactic mass centroids with degenerate neutrino object mass centroids will be expected to cause asymmetric galactic rotation curves. Asymmetric rotation curves mean that when observational rotational speeds from one side of a galaxy are folded over on top of the speeds on the other side, there are differences for the same galactic center distance. Some fraction of all galaxies should have asymmetric rotation curves. In this paper, we give one example where we fit the galaxy Hickson 88A (N6978)$^{24}$, within a condensed neutrino object. No attempt was made to optimize the fit at all.  The geometry of separated centroids is given in Figure 8. Here R$_{0}$ is the separation distance between the degenerate neutrino mass centroid and the galactic mass centroid, $\Theta_{0}$ is the phase angle between the embedded galaxy's rotation line and the vector $\vec{R}_{0}$, $\zeta$ is the one-sided farthest distance along the rotation line, $\sigma$ is the other sided distance and q$_{max}$, q$_{min}$ are the maximum and minimum distances from the neutrino matter centroid. The condensed neutrino model parameters used in Table 4 are x(0) = 0.05, $\Theta_{0} = 58^{\circ}$ and R$_{0}$ = 63.4 kpc, while the remaining parameters in Figure 8 are attributes of the embedded galaxy. In this example, we see clearly the asymmetric rotation curve of Hickson 88A due to the presence of the neutrino object's gradient gravitational potential.

\begin{table}[h]
\begin{center}
     \begin{tabular}{ccc}
     \multicolumn{3}{c}{ $m_{\nu}c^{2}$ = 1 eV} \\  \hline
       distance (kpc) & model  (km/s) & data (km/s) \\ \hline
       \mbox{} & \mbox{} & \mbox{} \\
    -15.03  &   -245 &  -230  \\
    -12.02 & -206 & -225 \\
    -9.01 & -161 & -220 \\
   -6.01  & -111 & -200 \\
  -3.00 & -58  & -110 \\
    0  & 0 & 0 \\
 3.00 & 61.5  & 115 \\
  6.01 & 126 & 225 \\
 9.01 & 194 & 260 \\
12.02 & 265 & 290  \\
15.03 & 338  & 320
      \end{tabular}
\end{center}
      \caption{Non-optimized model for Hickson 88A rotational asymmetry}
      \end{table}

\begin{figure}[htp]
\includegraphics[scale=0.75]{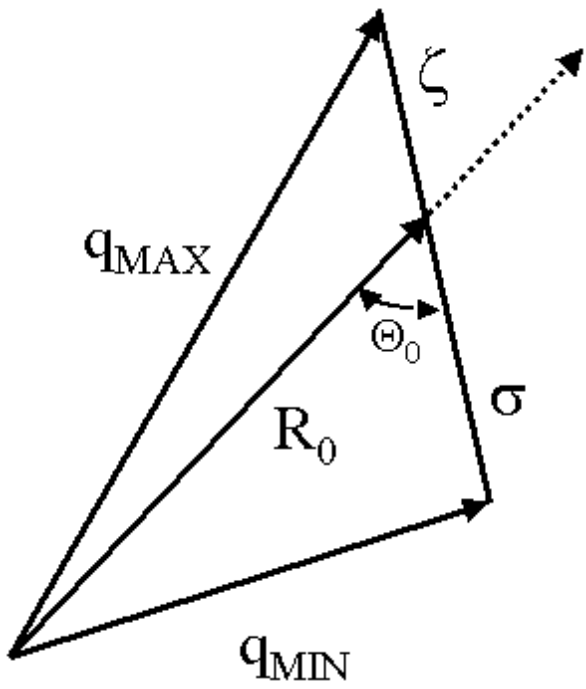}
\caption[*]{Geometry of separated centroids.}
\label{Fig. 8}
\end{figure}

\section{Conclusion}
New calculations are presented for degenerate condensed neutrino matter. Though transparent to light, condensed neutrino matter has several calculable effects on light, which are experimentally available for detection: embedded galaxies experience additional red-shifts from the work that light must do to overcome the neutrino gravitational well. Additionally, the neutrino well has calculable gravitational lensing effects.  This topic will be addressed in a future paper. If condensed neutrino matter exists, then SHM of embedded galaxies is an unavoidable occurrence.  Embedded galaxies that are not coincident with the centroid of a neutrino object can have asymmetric rotation profiles. An example, given in this paper, is Hickson 88A. The `billiard ball' configuration of multiple neutrino objects is unavoidably a result of their Fermi repulsion and it would lead to weird empirical lensing solutions. Neutrino condensation removes the free-streaming neutrino problem.

\end{document}